\documentclass[journal]{IEEEtran} 
\usepackage[dvipdf]{graphicx}
\graphicspath{}
\usepackage{mathtools}
\usepackage{amssymb}
\usepackage{multirow}
\usepackage{subfigure}
\usepackage[usenames,dvipsnames,svgnames,table]{xcolor}
\usepackage{cite}
\usepackage{soul}
\usepackage[percent]{overpic}
\usepackage{accents}
\usepackage{enumitem}   

\newcommand{\overbar}[1]{\mkern 1.5mu\overline{\mkern-1.5mu#1\mkern-1.5mu}\mkern 1.5mu}
\newcommand{\underlineb}[1]{\mkern 1.5mu\underline{\mkern-1.5mu#1\mkern-1.5mu}\mkern 1.5mu}

\begin{document}
%
\title{\vspace{-1cm}\LARGE{Model Predictive Control for Distributed Microgrid\\Battery Energy Storage Systems}}
%
%
%
\author{Thomas~Morstyn,~\IEEEmembership{Member,~IEEE,}
        Branislav~Hredzak,~\IEEEmembership{Senior Member,~IEEE,}\\
				Ricardo~P.~Aguilera,~\IEEEmembership{Member,~IEEE,} and
        Vassilios~G.~Agelidis,~\IEEEmembership{Fellow,~IEEE}
\thanks{T. Morstyn is with the Department of Engineering Science at the University of Oxford, Oxford OX1 2JD, United Kingdom (email: thomas.morstyn@eng.ox.ac.uk)}
\thanks{B. Hredzak is with the School of Electrical Engineering and Telecommunications at the University of New South Wales (UNSW Australia), Sydney, NSW 2052, Australia (email: b.hredzak@unsw.edu.au).}
\thanks{R. P. Aguilera is with The School of Electrical, Mechanical and Mechatronic Systems, University of Technology Sydney (UTS), NSW 2007, Australia. (email: raguilera@ieee.org).}
\thanks{V. G. Agelidis is with the Department of Electrical Engineering at the Technical University of Denmark (DTU), 2800 Kgs. Lyngby, Denmark (email: vasagel@elektro.dtu.dk).}
}

\maketitle

\begin{abstract}
\textcolor{Black}{This paper proposes a new convex model predictive control strategy for dynamic optimal power flow between battery energy storage systems distributed in an AC microgrid. The proposed control strategy uses a new problem formulation, based on a linear d--q reference frame voltage-current model and linearised power flow approximations. This allows the optimal power flows to be solved as a convex optimisation problem, for which fast and robust solvers exist. The proposed method does not assume real and reactive power flows are decoupled, allowing line losses, voltage constraints and converter current constraints to be addressed. In addition, non-linear variations in the charge and discharge efficiencies of lithium ion batteries are analysed and included in the control strategy. Real-time digital simulations were carried out for an islanded microgrid based on the IEEE 13 bus prototypical feeder, with distributed battery energy storage systems and intermittent photovoltaic generation. It is shown that the proposed control strategy approaches the performance of a strategy based on non-convex optimisation, while reducing the required computation time by a factor of 1000, making it suitable for a real-time model predictive control implementation.}

\end{abstract}

\begin{IEEEkeywords}
Battery Energy Storage, Energy Management, Microgrid, Model Predictive Control, Optimal Power Flow, Quadratic Programming.
\end{IEEEkeywords}

\IEEEpeerreviewmaketitle

\section{Introduction}
\IEEEPARstart{P}{ower} networks are undergoing a shift from the traditional model of centralised power generation towards a smart decentralised generation model. This trend encompasses the introduction of distributed renewable generation sources, controllable energy storage (ES) devices and advanced control using modern communication technologies \cite{Nehrir2011}. 

Microgrids have been proposed as an organising principle for this future smart decentralised grid \cite{Lasseter2002}. A microgrid is a collocated set of generation sources, loads and storage devices, which can operate as part of the main grid, or autonomously if islanded.

Power network control involves dispatching sources based on a wide range of competing objectives and constraints. Power network objectives include power loss minimisation and maximising the utilisation of renewable sources. Constraints are introduced by power quality limits and device operating limits. The solution of this constrained optimisation problem for static networks has been widely studied as the optimal power flow problem \cite{Momoh1999}.

The introduction of ES systems significantly increases the computational complexity of the optimal power flow problem, since it must be solved across a time horizon as the dynamic optimal power flow problem \cite{Levron2013}. The optimal use of ES systems depends not only on the network topology, constraints and objective, but also the state of charge (SoC) of the ES systems. SoC limits are of particular importance for battery ES systems, since they suffer significant lifetime deterioration when overcharged or undercharged \cite{Agarwal2010}. Also, batteries with very low SoC are unable to provide power to the network, due to their reduced output voltage \cite{Chen2006}. The SoC of the battery ES systems will also affect the optimisation objective function, since battery efficiency depends on the SoC \cite{Parvini2015}.

Recently, there has been significant research interest in using model predictive control (MPC) for microgrids with ES \cite{Olivares2014}. The network objective function and constraints are formulated into a finite-time optimal control problem. At each sampling period a set of system states are updated, the optimal control problem is solved online and the controller time horizon recedes by another step. \textcolor{Black}{MPC has been widely applied in the process industries, robotics and vehicle navigation \cite{Christofides2013,Lee2011}. With advances in communications and computer processing, it has become a practical solution for microgrid control.}

{\color{Black}

Existing MPC strategies for microgrids with distributed ES can be broadly divided into four groups:
\begin{enumerate}[label=(\roman*)]
\item Strategies which only consider a single ES system, or multiple ES systems in aggregate \cite{Parisio2014, Perez2013, Mayhorn2012}. This restricts power flows between different ES systems from being considered. 

\item Strategies assuming an ideal real power transfer model between the microgrid ES systems \cite{Dagdougui2014}, \cite{Garcia-Torres2015}. Losses and line power flow limits are not considered.

\item Strategies based on the DC power flow approximation \cite{Zeng2014}. In this case, lines connecting the ES systems are treated as purely reactive, so that with a small angle approximation power flows depend linearly on the bus voltage angles. This provides a convex optimisation for which fast and robust solvers are readily available. However, line losses and bus voltage limits are not considered. This is particularly unsuitable for microgrids, which may have relatively low line X/R ratios \cite{Tabatabaee2011}.

\item Strategies based on non-convex optimisation. MPC based on non-linear programming (NLP) in unbalanced microgrids is presented in \cite{Olivares2014b}. However, since the problem is non-convex, scalability is limited, and existing solvers are only guaranteed to find locally optimal solutions. An alternate approach is to use recursive dynamic programming for the ES system optimal power flow problem \cite{Levron2013}. This method obtains a globally optimal solution, but its numerical complexity grows in power law with the number of ES systems.
\end{enumerate}
}


{\color{Black}
This paper proposes a new convex MPC strategy for dynamic optimal flow between battery ES systems distributed in an AC microgrid. The novel features of the control strategy are:
\begin{enumerate}
\item A new formulation of the dynamic optimal power flow problem is proposed, based on a linear d--q reference frame voltage-current model and linearised power flow approximations. This allows the ES system optimal power flows to be solved as a convex quadratically constrained quadratic program (QCQP), for which fast and robust solvers exist, making it suitable for a real-time receding horizon MPC implementation. The proposed method does not assume real and reactive power flows are decoupled, allowing line losses, voltage constraints and converter current constraints to be addressed in the optimisation.
\item Non-linear variations in the charge and discharge efficiencies of lithium ion batteries are analysed with respect to SoC and output power. It is shown that this non-linear dependence is well approximated by second order polynomial functions, which are incorporated into the control strategy by updating the charge and discharge efficiencies at each sampling interval. 
\end{enumerate}}
To demonstrate the performance of the proposed control strategy, simulations were carried out for an islanded microgrid based on the IEEE 13 bus prototypical feeder, with distributed battery ES systems and intermittent PV generation. The simulations were completed with an RTDS Technologies real-time digital simulator, with non-linear battery models and switching converter models.  

The rest of this paper is organised as follows. Section \ref{sec:Principle} presents the principle of operation of the proposed MPC strategy. In Section \ref{sec:Modelling}, models are developed for the microgrid network and battery ES systems. Section \ref{sec:MPC} presents the MPC optimisation problem formulation. Results, demonstrating the control strategy's performance are presented in Section \ref{sec:Results}. Section \ref{sec:Conclusion} concludes the paper.

\section{Principle of Operation}
\label{sec:Principle}

This study considers a microgrid with distributed renewable generation sources and battery ES devices. The renewable generation sources and ES devices are connected to the microgrid by controllable voltage source converters (VSC). 

A standard method for VSC control is to have an inner loop current controller, and outer loop voltage controller which regulates the VSC output voltage to a desired reference \cite{Pogaku2007}. The VSC local control operates on a fast time-scale, and thus replacing it with a central MPC would be infeasible, requiring very high bandwidth communications.

Following a change in the VSC output voltage references, the time taken for a microgrid to reach steady state is on the order of 1 second. Assuming that the local controllers are well designed and give desirable transient performance, the MPC strategy can operate on a slower time-scale, with a static network model. 

\begin{figure}
\centering
\includegraphics[trim=0cm 0.0cm 8cm 23cm, clip=true, width=0.9\columnwidth]{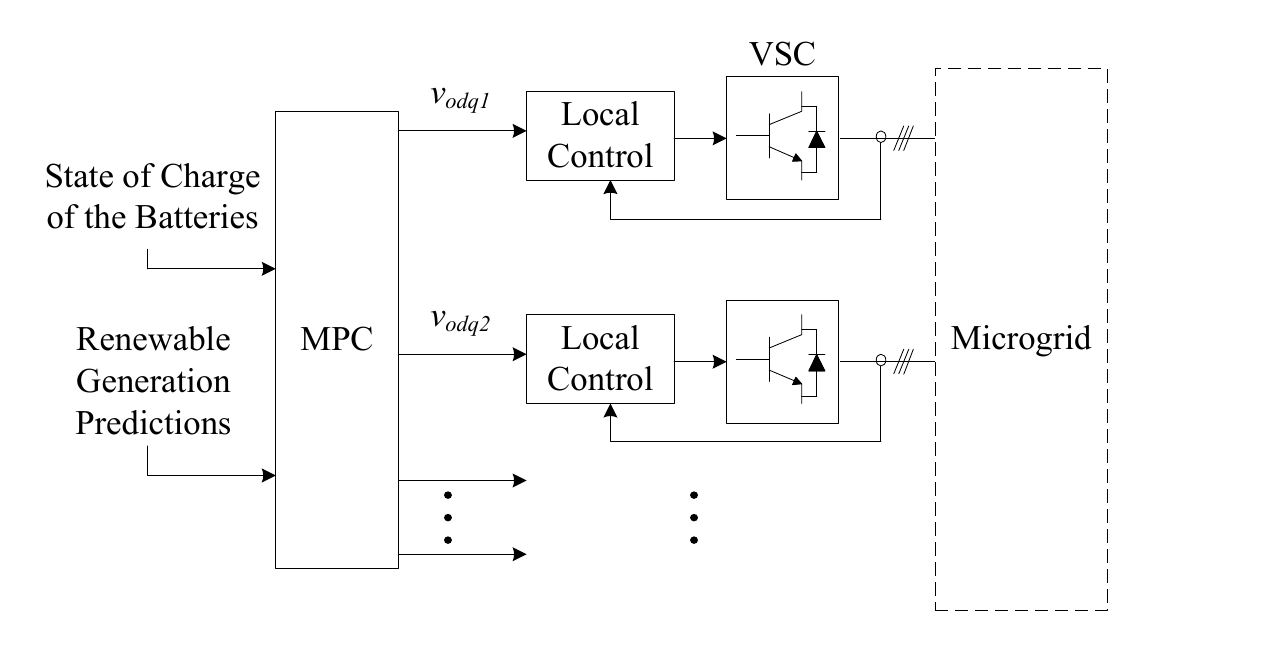}
\caption{High level block diagram of the proposed MPC strategy.}
\label{fig:HighLevelControl}
\end{figure}

In this study a 1 minute sampling period is chosen, allowing the power flows to be adjusted in response to changes in PV generation, taking into account the SoC of the batteries. The manipulated variables of the MPC are the output voltages of the converters, transformed to a fixed frequency synchronous d--q reference frame. These are supplied as references to the local VSC voltage controllers. A high level block diagram of the proposed control strategy is shown in Fig. \ref{fig:HighLevelControl}.


This study considers an islanded microgrid operating autonomously from the main grid. Since the microgrid load must be supplied by the local generation sources and ES systems, a natural objective is power network loss minimisation, while making use of the maximum power available from renewable generation. Network power quality requirements impose constraints on the VSC output voltages, and device operating limits require constraints on the VSC output currents and SoC of the batteries.

\section{Microgrid Modelling}
\label{sec:Modelling}


\subsection{Microgrid Network}

\begin{figure}[t!]
\centering
\includegraphics[trim=0cm 0.2cm 7cm 22.7cm, clip=true, width=1\columnwidth]{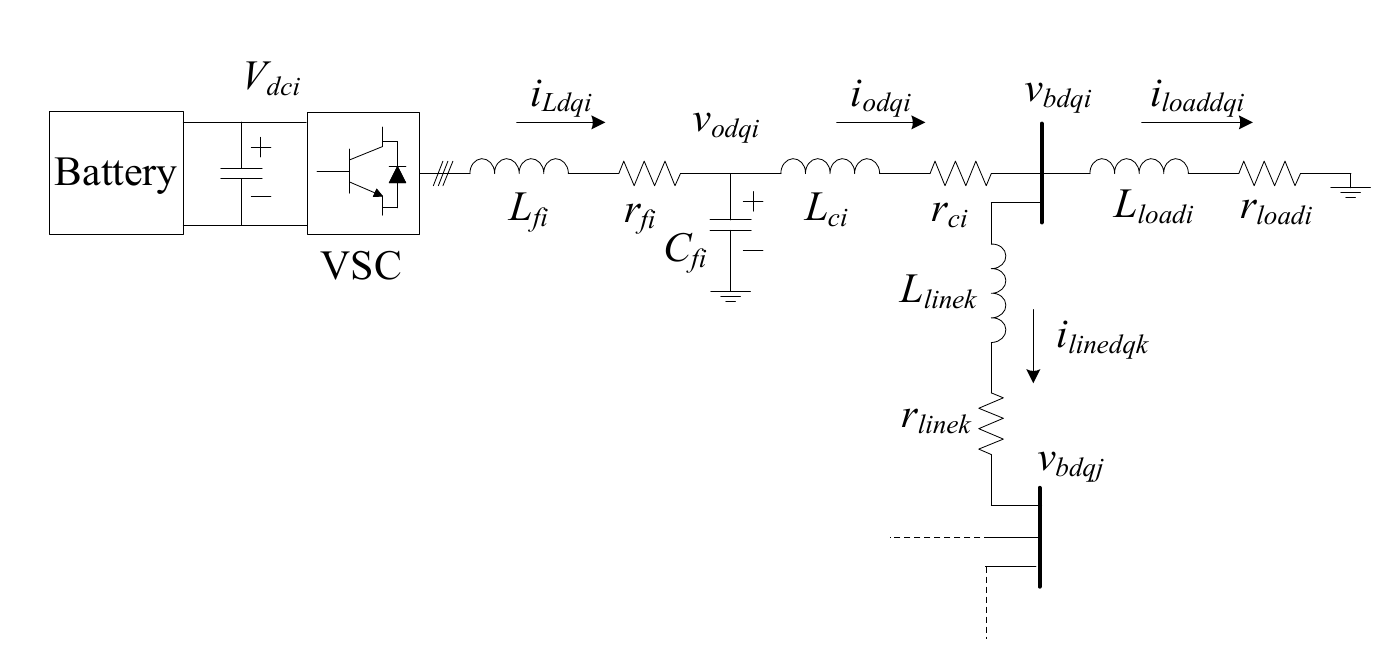}
\caption{Two bus microgrid segment, with a battery energy storage system, LCL output filter, RL line and RL load.}
\label{fig:MGsection}
\end{figure}

The network model considers the passive components, including coupling filters, lines and loads, between the buses which are actively regulated by the VSCs. 


Fig. \ref{fig:MGsection} shows a two bus segment of the microgrid. The full microgrid has loads $S_{loads}=\{1,\dots,N_{load}\}$, lines $S_{line}=\{1,\dots,N_{line}\}$ and VSCs $S_{vsc}=\{1,\dots,N_{vsc}\}$. The microgrid lines are modelled as RL circuits. In the synchronous d--q reference frame, the state equations for the VSC output currents, load currents and line currents can be combined into a state space model, with the vector of regulated VSC output voltages as the input \cite{Pogaku2007}.
\begin{align}
\frac{d}{dt}\begin{bmatrix}{\+i}_{odq}\\ {\+i}_{loaddq}\\ {\+i}_{linedq}\end{bmatrix} &=A_{net}\begin{bmatrix}\+i_{odq}\\ \+i_{loaddq}\\ \+i_{linedq}\end{bmatrix} +B_{net}\+v_{odq},\\
\+i_{odq} &= \begin{bmatrix}i_{odq1}^{T}&\cdots&i_{odqN_{vsc}}^{T}\end{bmatrix}^{T}, \nonumber\\ 
\+i_{loaddq} &= \begin{bmatrix}i_{loaddq1}^{T}&\cdots&i_{loaddqN_{load}}^{T}\end{bmatrix}^{T}, \nonumber\\
\+i_{linedq} &=  \begin{bmatrix}i_{linedq1}^{T}&\cdots&i_{linedqN_{line}}^{T}\end{bmatrix}^{T},  \nonumber\\
\+v_{odq} &=  \begin{bmatrix}v_{odq1}^{T}&\cdots&v_{odqN_{vsc}}^{T}\end{bmatrix}^{T}.  \nonumber
\end{align}

$i_{odq1}$ and $v_{odqi}$ are the d--q component vectors of the output current and output voltage of the $i$th VSC. The d--q current components of the load at bus $i$ are given by $i_{loaddqi}$, and the d--q current components of line $j$ are given by $i_{linedqj}$.

The MPC strategy operates on a much slower time-scale than the network dynamics. The following static model describing the steady state network behaviour is used to calculate the network currents from the VSC output voltages. 
\begin{align}
&\+i_{odq} = \+G_{i_o}\+v_{odq},~ {\+i}_{loaddq} = \+G_{i_{load}}\+v_{odq}\\ 
&\+i_{linedq}=\+G_{i_{line}}\+v_{odq},\nonumber\\
&[\+G_{i_o}^{T}~\+G_{i_{load}}^{T}~\+G_{i_{line}}^{T}]^{T}=(-A_{net})^{-1}B_{net},\nonumber \\
&\+G_{i_o} = [G_{i_{o1}}^{T}\cdots G_{i_{oN_{vsc}}}^{T}]^{T},~
\+G_{i_{load}} = [G_{i_{load1}}^{T}\cdots G_{i_{loadN_{load}}}^{T}]^{T},\nonumber\\
&\+G_{i_{line}}= [G_{i_{line1}}^{T}\cdots G_{i_{lineN_{line}}}^{T}]^{T}.\nonumber
\end{align}

\textcolor{Black}{The d--q reference frame voltage-current model is based on a balanced three phase system. In general, this assumption is justified by power network standards which place stringent restrictions on the level of voltage unbalance. However, if voltage unbalance is a concern it can be addressed with a lower level compensation scheme based on virtual impedance \cite{Savaghebi2012}.}

\textcolor{Black}{The microgrid impedances are required to develop the voltage-current model. If the microgrid impedances are not available, they can be obtained using online identification methods \cite{Ciobotaru2007}.}

\subsection{Voltage Source Converters}
The MPC strategy controls power flows in the microgrid by providing voltage references to the VSCs of the renewable generation sources and ES systems. The VSCs use standard decoupled d--q inner loop current controllers, and outer loop voltage controllers to regulate their output voltage to the desired reference. 

Let $\omega$ be the microgrid frequency. In the synchronous d--q reference frame, the state equations for inductor current and output voltages of VSC $i\in S_{vsc}$ are given by \cite{Wu2007},
\begin{align}
\frac{d}{dt}i_{Ldi} &= -\frac{r_{fi}}{L_{fi}}i_{Ldi} + \omega i_{Lqi} -\frac{1}{L_{fi}}v_{odi} + \frac{a_{mi} V_{dci}}{L_{fi}}{u}_{di},\\
\frac{d}{dt}i_{Lqi} &= -\frac{r_{fi}}{L_{fi}}i_{Lqi} - \omega i_{Ldi} -\frac{1}{L_{fi}}v_{oqi} + \frac{a_{mi} V_{dci}}{L_{fi}}{u}_{qi},\nonumber\\
\frac{d}{dt}v_{odi} &= \frac{1}{C_{fi}}(i_{Ldi}-i_{odi})+\omega v_{oqi},\nonumber \\
\frac{d}{dt}v_{oqi} &= \frac{1}{C_{fi}}(i_{Lqi}-i_{oqi})-\omega v_{odi}. \nonumber
\end{align}
$V_{dci}$ is the VSC DC link voltage, $a_{mi}=0.5$ for sinusoidal pulse width modulation (PWM), $i_{Ldqi}$ is the inductor current d--q component vector and $u_{dqi}$ is the PWM control signal d--q component vector. 

The local VSC controllers regulate the output voltage much faster than the MPC sampling period. Therefore, the following static VSC model can be used. 
\begin{align}
i_{Ldqi} &= G_{i_{Li}}\+v_{odq}\\
\tilde{u}_{dqi}&=G_{\tilde{u}_{i}}\+v_{odq},~
\tilde{u}_{dqi} = a_{mi}V_{dci}[u_{di}~u_{qi}]^{T}\nonumber\\
G_{i_{Li}} & =\begin{bmatrix}0&-\omega C_{fi}\\\omega C_{fi}&0 \end{bmatrix}\bigg(e_{i}^{T}\otimes\begin{bmatrix}1&0\\0&1\end{bmatrix}\bigg)+G_{i_{oi}}\nonumber\\
G_{\tilde{u}_{i}}&=\begin{bmatrix}r_{fi}&-\omega L_{fi}\nonumber\\\omega L_{fi}&r_{fi} \end{bmatrix} G_{i_{Li}}+\bigg(e_{i}^{T}\otimes\begin{bmatrix}1&0\\0&1\end{bmatrix}\bigg)\nonumber
\end{align}
$e_i$ is a vector of length $N_{vsc}$, with a 1 at position $i$ and zeros for all other entries. $\tilde{u}_{dqi}$ is the VSC input voltage d--q component vector. The real power supplied from the DC side of the VSC is given by
\begin{align}
\label{equ:Pvsc}
P_{vsci} = \tilde{u}_{di}i_{Ldi} + \tilde{u}_{qi}i_{Lqi}.
\end{align}

\subsection{Battery Energy Storage System}
\label{sec:BESmodel}

Let the subset of microgrid VSCs associated with battery ES systems be given by $S_{batt}\subseteq S_{vsc}$. The SoC dynamics for battery ES system $i\in S_{batt}$ can be modelled using the following discrete time state equations \cite{Zeng2014}, \cite{Olivares2014b},
\begin{align}
\label{equ:SOCmodel}
SoC_{i}(k+1) = \begin{cases}
                        SoC_{i}(k) - \eta_{chi}(k)T_{s}\frac{ P_{vsci}(k)}{E_{maxi}},~ P_{vsci}(k)\leq0, \\
                        SoC_{i}(k) - \frac{1}{\eta_{disi}(k)}T_{s}\frac{ P_{vsci}(k)}{E_{maxi}},~ P_{vsci}(k)>0.
                    \end{cases}							
\end{align}
$T_{s}$ is the sampling period, $E_{maxi}$ is the maximum battery energy in Ws, $\eta_{chi}$ is the charging efficiency and $\eta_{disi}$ is the discharging efficiency.

When this model is used for MPC the charge and discharge efficiencies are commonly assumed to be constant \cite{Zeng2014}, \cite{Olivares2014b}. However, the battery efficiency is actually a non-linear function of the SoC and the battery output power.  

At a particular steady state output current, an electrochemical battery cell (e.g. lithium ion, lead acid) can be modelled as an open circuit voltage $V_{oci}$ and terminal resistance $R_{ti}$, each with a non-linear dependence on the SoC \cite{Chen2006}.
\begin{align}
\label{equ:cellmodel}
V_{oci} &= a_{0i}e^{-a_{1i}SoC_{i}}+a_{2i}+a_{3i}SoC_{i}-a_{4i}SoC_{i}^{2}+a_{5i}SoC_{i}^{3},\nonumber\\
R_{ti}&=R_{si}+R_{tsi}+R_{tli}\nonumber\\
R_{si} &= b_{0i}e^{-b_{1i}SoC_{i}}+b_{2i}+b_{3i}SoC_{i}-b_{4i}SoC_{i}^{2}+b_{5i}SoC_{i}^{3}, \nonumber\\
R_{tsi} &= c_{0i}e^{-c_{1i}SoC}+c_{2i},~R_{tli} = d_{0i}e^{-d_{1i}SoC_{i}}+d_{2i}.
\end{align}

Assuming there is a balancing circuit between the battery cells, the total battery output current $i_{dci} = i_{celli}N_{parai}$ and the battery output voltage $V_{dci}=v_{celli}N_{seriesi}$, where $N_{seriesi}$ is the number of series connected cells in each string and $N_{parai}$ is the number of parallel connected cell strings.

\textcolor{Black}{For a particular SoC and output power, the battery cell current can be obtained by solving the following quadratic equation,}
\begin{align}
\textcolor{Black}{i_{celli}^{2}-i_{celli}\frac{V_{oci}(SoC_{i})}{R_{ti}(SoC_{i})}+\frac{1}{N_{seriesi}N_{parai}}\frac{P_{vsci}}{R_{ti}(SoC_{i})}=0.}
\end{align}
\textcolor{Black}{Then, the battery charging and discharging efficiencies are given by,}
\begin{align}
\textcolor{Black}{\eta_{chi} = \frac{i_{celli}V_{oci}(SoC_{i})}{i_{celli}(V_{oci}(SoC_{i})-i_{celli}R_{ti}(SoC_{i}))},~i_{celli}\leq0} \\
\textcolor{Black}{\eta_{disi} = \frac{i_{celli}(V_{oci}(SoC_{i})-i_{celli}R_{ti}(SoC_{i}))}{i_{celli}V_{oci}(SoC_{i})},~i_{celli}>0.\nonumber}
\end{align}   

The microgrid considered in this study includes 100kWh, 900V lithium ion batteries. The batteries are made up of 130 parallel strings, each with 215 series connected 4.2V, 860mAh cells. The cell model parameters in (\ref{equ:cellmodel}) are provided by \cite{Kim2011}. 

\begin{figure}
\centering
\includegraphics[width=0.8\columnwidth,trim={0 0 9cm 21cm},clip]{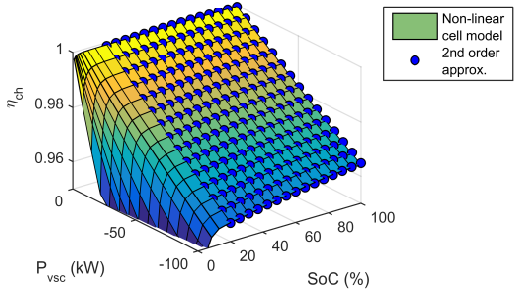}
\caption{Relationship between VSC output power and battery SoC with $\eta_{ch}$, and a fitted second order polynomial approximation.}
\label{fig:Eff_Charge}
\end{figure}

\begin{figure}
\centering
\includegraphics[width=0.8\columnwidth,trim={0 0 9cm 21cm},clip]{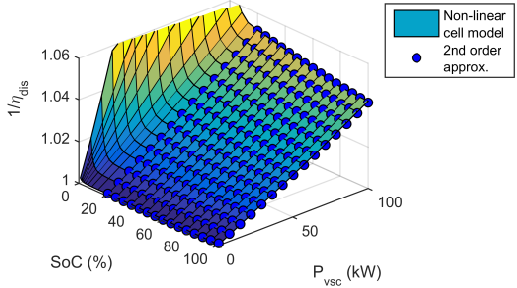}
\caption{Relationship between VSC output power and battery SoC with $\frac{1}{\eta_{dis}}$, and a fitted second order polynomial approximation.}
\label{fig:Eff_Discharge}
\end{figure}

Fig. \ref{fig:Eff_Charge} and \ref{fig:Eff_Discharge} show the effect of the SoC and output power on the charge and discharge efficiency. As shown, second order polynomials provide good approximations for the efficiency over the ranges of interest for the output power (-100kW to 100kW) and SoC (20\% to 100\%). 
\begin{align}
\label{equ:EffAprox}
\eta_{chi} = p_{0i}^{ch}+p_{SoC1i}^{ch}SoC_{i}+p_{SoC2i}^{ch}SoC_{i}^{2}+p_{P1i}^{ch}P_{vsci}\nonumber\\+p_{P2i}^{ch}P_{vsci}^{2}+p_{PSoCi}^{ch}SoC_{i}P_{vsci}, \\
\frac{1}{\eta_{disi}} = p_{0i}^{dis}+p_{SoC1i}^{dis}SoC_{i}+p_{SoC2i}^{dis}SoC^{2}_{i}+p_{P1i}^{dis}P_{vsci} \nonumber\\+p_{P2i}^{dis}P_{vsci}^{2}+p_{PSoCi}^{dis}SoC_{i}P_{vsci}.\nonumber
\end{align}

\section{Model Predictive Control Formulation}
\label{sec:MPC}

In this section the microgrid optimisation problem is formulated as a convex QCQP, which can be implemented with a receding horizon for MPC. 

Let $\tau =\{k_{0},\dots,k_{0}+N_{p}-1\}$ be the MPC time horizon. The MPC state variables are the battery ES systems' SoC levels $SoC_i(k),k\in\tau~i\in S_{batt}$. The MPC manipulated variables are the VSC d--q reference frame output voltage references, VSC charging powers and VSC discharging powers, over the time horizon.
\begin{align}
\hat{U} &= [U(k_{0})^{T}~U(k_{0}+1)^{T}\cdots U(k_{0}+N_{p}-1)^{T}]^{T},\\
&U(k) = [\+v_{odq}(k)^{T}~\+P_{ch}^{T}(k)~\+P_{dis}^{T}(k)]^{T}, \nonumber\\
&\+P_{ch}(k) = [P_{ch1}(k)\cdots P_{chN_{vsc}}(k)]^{T},\nonumber\\
&\+P_{dis}(k) = [P_{dis1}(k)\cdots P_{disN_{vsc}}(k)]^{T}. \nonumber
\end{align}
The charge and discharge power variables are introduced to allow different charge and discharge efficiencies while maintaining a convex QP formulation.

\subsection{Objective Function}

The objective chosen for the islanded microgrid is to minimise real power losses over the time horizon. Real power losses are incurred due to network line resistances, resistances in the VSC LCL filters and losses from charging/discharging the batteries. The objective function to be minimised is given by,
\begin{align}
\label{equ:Obj}
J_{obj} &= \sum_{k\in\tau}J_{LCL}(k)+J_{line}(k)+J_{batt}(k),\\
J_{LCL}(k) &= \sum_{i\in S_{vsc}}\+v_{odq}^{T}(k)\big{(}r_{fi}G_{i_{Li}}^{T}(k)G_{i_{Li}}(k)\nonumber\\
&~~~~~~~~+r_{ci}G_{i_{oi}}^{T}(k)G_{i_{oi}}(k)\big{)}\+v_{odq}(k),\\
J_{line}(k) &= \sum_{j\in S_{line}}r_{linej}\+v_{odq}^{T}(k)G_{i_{linej}}^{T}G_{i_{linej}}\+v_{odq}(k),\\
J_{batt}(k) &=\sum_{i\in S_{batt}}(1-\eta_{chi})P_{chi}(k)+\left(\frac{1}{\eta_{disi}}-1\right)P_{disi}(k).
\end{align}



\textcolor{Black}{The losses associated with the relative positions of the microgrid loads, renewable generation sources and battery ES systems are included in the objective function through the $G_{i_{oi}}(k)$, $G_{i_{Li}}(k)$ and $G_{i_{linej}}$ matrices. $G_{i_{oi}}(k)$ and $G_{i_{Li}}(k)$ vary over the time horizon $k\in\tau$ based on the microgrid load predictions.}

Directly including the SoC and output power dependence of $\eta_{chi}$ and $\eta_{disi}$ in the objective function would make the optimisation problem non-convex. Instead, the efficiencies are updated each time interval prior to the optimisation, and treated as constant over the time horizon. The efficiencies are calculated based on the current SoC estimates and the output powers from the previous time interval, using the second order polynomial approximations (\ref{equ:EffAprox}).




\subsection{Constraints}

The microgrid power quality requirements and device operating limits are formulated as affine equality and convex inequality constraints to provide a convex QCQP formulation.

\subsubsection{VSC Output Power}

To ensure the MPC solution is feasible, the VSC charge and discharge power manipulated variables must match the power flows expected from the output voltage manipulated variables. 
\begin{align}
\label{equ:PconstQ}
P_{disi}(k)-&P_{chi}(k) = \tilde{u}_{di}(k)i_{Ldi}(k) + \tilde{u}_{qi}(k)i_{Lqi}(k),\\
\label{equ:Pconst0}
&0\leq P_{chi}(k),~0\leq P_{disi}(k),~i\in S_{vsc}~k\in\tau.
\end{align}
\textcolor{Black}{The quadratic equality constraint (\ref{equ:PconstQ}) makes the microgrid optimisation problem non-convex. To obtain a convex formulation, the quadratic equality constraints are replaced by approximate affine equality constraints obtained through linearisation, based on nominal operating points of $I_{Ldqi}(k),~\tilde{U}_{dqi}(k)$, for the VSC inductor currents and input voltages.}
\begin{align}
\label{equ:PconstL}
P_{disi}(k)-&P_{chi}(k) 
= [\tilde{U}_{di}(k)~\tilde{U}_{qi}(k)]G_{i_{Li}}(k)\+v_{odq}(k)\\
&+[I_{Ldi}(k)~I_{Lqi}(k)]G_{\tilde{u}_{i}}(k)\+v_{odq}(k)\nonumber
\\&-\tilde{U}_{di}(k)I_{Ldi}(k)-\tilde{U}_{qi}(k)I_{Lqi}(k),~i\in S_{vsc}~k\in\tau.\nonumber
\end{align}

For the $i$th battery ES system, let the allowed range of real output powers be given by $[-\bar{P}_{chi},~\bar{P}_{disi}]$. This motivates the following constraints,
\begin{align}
\label{equ:Pbattconst}
P_{chi}(k)\leq \bar{P}_{chi},~P_{disi}(k)\leq \bar{P}_{disi},~i\in S_{batt}~k\in\tau.
\end{align}

\textcolor{Black}{The remaining VSCs ($i\in S_{vsc}$, $i\notin S_{batt}$) are associated with renewable generation sources or constant power loads. Let $S_{gen}\subseteq S_{vsc}$ be the set of renewable generation sources and $S_{cpl}\subseteq S_{vsc}$ be the set of constant power loads.}

\textcolor{Black}{The constant power load predictions over the time horizon are given by $P_{cpli}(k),~i\in S_{cpl}~k\in\tau$. The constant power loads are constrained to operate at this level.}
\begin{align}
\label{equ:Pcplconst}
\textcolor{Black}{
P_{disi}(k) =0,~P_{chi}(k) = P_{cpli}(k),~i\in S_{cpl}~k\in\tau.}
\end{align}

The available renewable generation predictions over the time horizon are given by $P_{mppi}(k),~i\in S_{gen}~k\in\tau$. The renewable generation sources are constrained to operate at this level.
\begin{align}
\label{equ:Pgenconst}
P_{disi}(k) =  P_{mppi}(k),~P_{chi}(k) = 0,~i\in S_{gen}~k\in\tau.
\end{align}

\textcolor{Black}{In practice, the linearisation (\ref{equ:PconstL}) may result in a mismatch between the available renewable generation $P_{mppi}(k)$ and the power required to achieve the output voltage reference $v_{odqi}(k)$. This is corrected by including an integral control loop with the local VSC voltage controller, which adjusts the $v_{odi}(k_0)$ reference until $P_{vsci}(k_0)=P_{mppi}(k_0)$.}

\subsubsection{Battery SoC}
A standard selection for the battery ES system SoC constraints is $\underlineb{SoC}=20\%$, $\overbar{SoC}=100\%$.
\begin{align}
\label{equ:SOCconst}
\underlineb{SoC}\leq SoC_{i}(k+1)\leq \overbar{SoC},i\in S_{batt}~k\in\tau\end{align}
Using (\ref{equ:SOCmodel}), the SoC at each interval of the time horizon can be expressed in terms of the estimates of the current SoC ($SoC_{i}(k_{0}),~i\in S_{batt}$) and the manipulated variables. 
\begin{align}
\label{equ:SOCmodelconst}
SoC_{i}(k+1) = SoC_{i}(k) + \eta_{chi}T_{s}\frac{P_{chi}(k)}{E_{maxi}}-\frac{1}{\eta_{disi}}T_{s}\frac{P_{disi}(k)}{E_{maxi}}.
\end{align}
Since a linear system model is necessary to maintain a convex QP formulation, $\eta_{chi}$ and $\eta_{disi}$ are updated each time interval prior to the optimisation, and treated as constant over the time horizon. 

\textcolor{Black}{The SoC model and objective function require that for each $k\in \tau$ and $i\in S_{batt}$ either $P_{chi}(k)=0$ or $P_{disi}(k)=0$. This is ensured by the constraint that the charging and discharging powers are greater than or equal to zero (\ref{equ:Pconst0}) and the observation that $J_{batt}(k)$ applies a cost to $P_{chi}(k)>0$ and $P_{disi}(k)>0$ since the batteries have limited efficiency \cite{Olivares2014b}. If the optimisation problem has a feasible solution with $P_{dis}(k)>0$ and $P_{chi}(k)>0$, there will be a lower cost solution with one at a lower value, and the other at zero.}


\subsubsection{VSC Output Voltage}

\begin{figure}
\centering
\includegraphics[width=0.8\columnwidth,trim={0 0 8.5cm 22.5cm},clip]{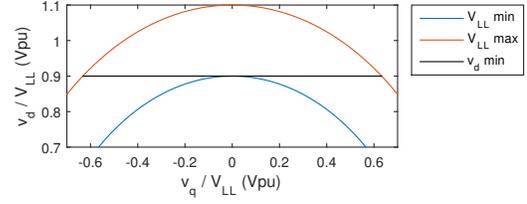}
\caption{VSC RMS output voltage constraints and conservative affine lower bound.}
\label{fig:Vconst}
\end{figure}

Standard RMS voltage limits for an AC microgrid are $V_{LL}\pm10\%$, where $V_{LL}$ is the nominal line to line voltage. Therefore, the VSC d--q  voltage components should be limited so that,
\begin{align}
\label{equ:VlimQU}
v_{odi}^{2}(k) + v_{oqi}^{2}(k) &\leq (1.1V_{LL})^{2}\\
\label{equ:VlimQL}
v_{odi}^{2}(k) + v_{oqi}^{2}(k) &\geq (0.9V_{LL})^{2},~i\in S_{vsc}~k\in\tau.
\end{align} 
\textcolor{Black}{The upper bound (\ref{equ:VlimQU}) is a convex quadratic constraint and can be directly included in the convex MPC formulation. However, the lower bound (\ref{equ:VlimQL}) is non-convex.} Assuming the voltage should be able to vary in the d--q reference frame around a nominal operating point of $v_{odqi}=(V_{LL}$,~0), the non-convex quadratic lower bound can be approximated by the following conservative affine inequality constraint on the d axis voltage. 
\begin{align}
\label{equ:VlimL}
 v_{odi}(k) \geq 0.9{V}_{LL},~i\in S_{vsc}~k\in\tau.
\end{align} 
Fig. \ref{fig:Vconst} shows the RMS voltage limits in the d--q reference frame and the conservative d axis lower bound. 

\subsubsection{VSC Output Current}

The maximum rated current of the VSC switches will impose a constraint on the VSC RMS current. Note that this implicitly introduces a limit on the maximum reactive output power of the VSC, based on the output voltage and real output power \cite{Johansson2004}. Let the maximum VSC RMS phase current be given by $\bar{I}_{phi}$. The VSC d--q inductor current components should be limited so that, 
\begin{align}
\label{equ:IlimQ}
i_{Ldi}^{2}(k) + i_{Lqi}^{2}(k) \leq (\sqrt{3}\bar{I}_{phi}(k))^{2},~i\in S_{vsc}~k\in\tau.
\end{align} 

\subsection{Non-Convex and Convex Optimisation Formulations}

The MPC loss minimisation objective is a quadratic function of the manipulated variables. The microgrid MPC strategy can be formulated as a non-convex QCQP given by,
\begin{align}
\label{equ:nonconvexQP}
\underset{\hat{U}}{\text{minimise}}~&J_{obj}\\
\text{subject~to~} &(\ref{equ:PconstQ}),(\ref{equ:Pconst0}),(\ref{equ:Pbattconst}),(\ref{equ:Pcplconst}),(\ref{equ:Pgenconst}),\nonumber\\
&(\ref{equ:SOCconst}),(\ref{equ:SOCmodelconst}),(\ref{equ:VlimQU}),(\ref{equ:VlimQL}),(\ref{equ:IlimQ}).\nonumber
\end{align}

\textcolor{Black}{Since the objective function is convex, the MPC optimisation can be formulated as a convex QCQP by replacing the VSC output power quadratic equality constraints (\ref{equ:PconstQ}) with affine approximations (\ref{equ:PconstL}), and by replacing the quadratic lower bound on the VSC RMS output voltages (\ref{equ:VlimQL}) with the conservative affine lower bound (\ref{equ:VlimL}).}
\begin{align}
\label{equ:convexQP}
\textcolor{Black}{\underset{\hat{U}}{\text{minimise}}}~&\textcolor{Black}{J_{obj}}\\
\textcolor{Black}{\text{subject~to~}} &\textcolor{Black}{(\ref{equ:Pconst0}),(\ref{equ:PconstL}),(\ref{equ:Pbattconst}),(\ref{equ:Pcplconst}),(\ref{equ:Pgenconst}),}\nonumber\\
&\textcolor{Black}{(\ref{equ:SOCconst}),(\ref{equ:SOCmodelconst}),(\ref{equ:VlimQU}),(\ref{equ:VlimL}),(\ref{equ:IlimQ}).}\nonumber
\end{align}

\textcolor{Black}{Having formulated the dynamic optimal power flow problem as a convex QCQP, it can be solved with interior point methods which can solve problems with hundreds of variables and constraints in several seconds on standard computing hardware \cite{Boyd2004}.}

\subsection{Model Predictive Control Implementation}
{\color{Black}
The convex QCQP is implemented using receding horizon MPC to control the microgrid battery ES systems in real-time. At each time interval, $k_{0}$: 
\begin{enumerate}
\item Updated battery SoC estimates, $SoC_i(k_{0}),~i\in S_{batt}$, renewable generation predictions, $P_{mppi}(k),~i\in S_{gen}~k\in\tau$, load predictions, $P_{cpli}(k),~i\in S_{cpl}~k\in\tau$, and network matricies $G_{i_{oi}}(k), G_{i_{Li}}(k),~i\in S_{vsc}~k\in\tau$ are obtained.

\item The convex dynamic optimal power flow problem (\ref{equ:convexQP}) is solved.
  
\item The optimal d--q voltages for the current time interval $v^{*}_{odqi}(k_{0}),i\in S_{vsc}$, are supplied as references for the local VSC voltage controllers. In addition, the renewable sources operated for maximum power point tracking are provided with, $P_{mppi}(k_{0}),~i\in S_{gen}$.

\item The MPC time horizon recedes by a step, $k_{0}\leftarrow k_{0} +1$, for the next time interval.
\end{enumerate}
}

\section{Results}
\label{sec:Results}

\begin{figure}[!t]
\includegraphics[trim=0cm 0.2cm 11.5cm 22cm, clip=true, width=0.8\columnwidth]{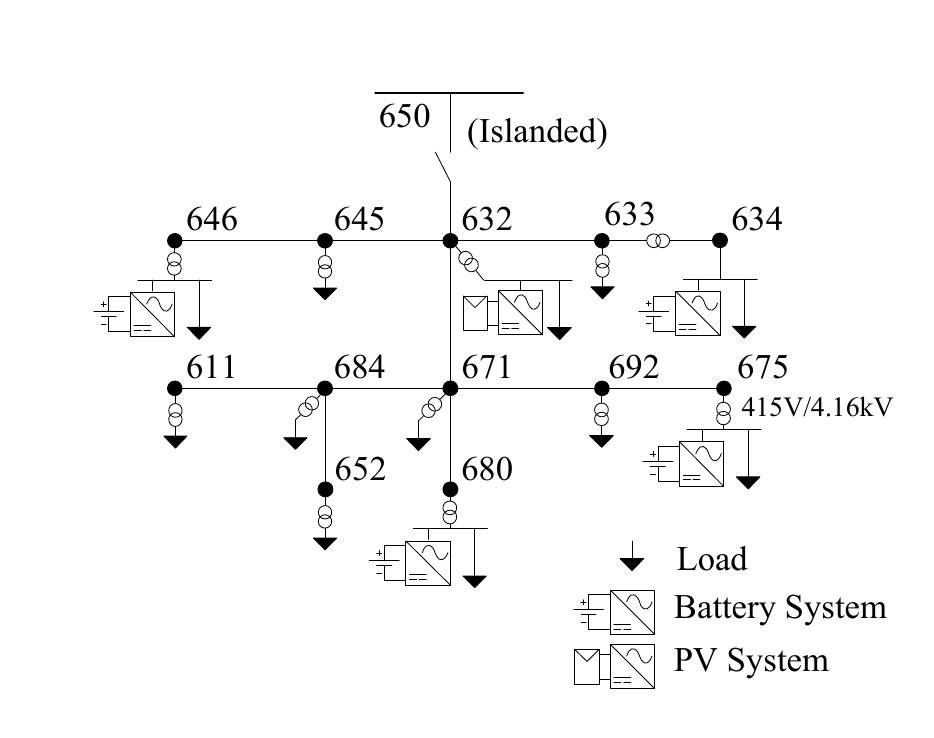}
\caption{Case study microgrid.}
\label{fig:IEEE13}
\end{figure}

To demonstrate the performance of the proposed MPC strategy, simulations were carried out using an RTDS Technologies real-time digital simulator. The proposed MPC strategy was used to control four battery ES systems and a PV generation source, distributed in an islanded microgrid based on the IEEE 13 bus prototypical test feeder. The batteries were simulated using the non-linear model from \cite{Kim2011}, and switching converter models were used for the VSCs.


\subsection{Real-Time Digital Simulation}
The case study microgrid is shown in Fig. \ref{fig:IEEE13}. The simulation parameters are provided in Table \ref{tab:param}. The network is operated at 50Hz. Each battery ES system has a 100kWh, 900V lithium ion battery, as described in Section \ref{sec:BESmodel}. PV generation with 100kW nominal capacity at standard test conditions is introduced at bus 632. The maximum PV power point was calculated using the method from \cite{Villalva2009} for irradiance and temperature data with 1 minute resolution, taken from the NREL Baseline Measurement Station in Colorado, for the 15th of August 2015 between 6am and 4pm. 


\begin{table}[!t]
\renewcommand{\arraystretch}{1.0}
\caption{Real-Time Simulation Parameters}\vspace{-0.5cm}
\label{tab:param}
{\small
\begin{tabular}[t]{llllll}
\hline
$\omega$&50Hz&$V_{LL}$&415V&$f_{vsc}$&10kHz\\
$N_{p}$&30min&$T_{s}$&1min&$\bar{I}_{ph}$&150A\\
$\underlineb{SoC}$&20$\%$&$\overbar{SoC}$&100$\%$&$\bar{P}_{ch}$&100kW\\
$\bar{P}_{dis}$&100kW&$L_{f}$&3.8mH&$r_{f}$&0.15$\Omega$\\
$C_{f}$&680$\mu$F&$L_{c}$&300$\mu$H&$r_{c}$&0.05$\Omega$\\
$p_{0}^{ch}$&1.00&$p_{SoC1}^{ch}$&4.00$\cdot$10$^{-3}$&$p_{SoC2}^{ch}$&-3.11$\cdot$10$^{-3}$\\
$p_{P1}^{ch}$&-4.77$\cdot$10$^{-7}$&$p_{P2}^{ch}$&3.06$\cdot$10$^{-13}$&$p_{PSoC}^{ch}$&9.66$\cdot$10$^{-8}$\\
$p_{0}^{dis}$&1.00&$p_{SoC1}^{dis}$&-4.60$\cdot$10$^{-3}$&$p_{SoC2}^{dis}$&4.13$\cdot$10$^{-3}$\\
$p_{P1}^{dis}$&5.00$\cdot$10$^{-7}$&$p_{P2}^{dis}$&4.23$\cdot$10$^{-13}$&$p_{PSoC}^{dis}$&-1.36$\cdot$10$^{-7}$\\
\hline
\end{tabular}
}
\end{table}

\begin{figure}[!t]
\centering
\includegraphics[width=0.8\columnwidth,trim={0.0cm 0cm 6.0cm 21cm}]{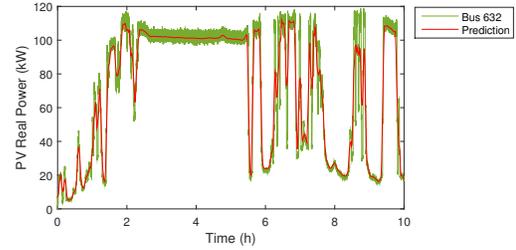}
\caption{\textcolor{Black}{PV generation real output power and predictions.}}
\label{fig:Ppv}
\end{figure}

\begin{figure}[!t]
\centering
\includegraphics[width=0.8\columnwidth,trim={0.0cm 0cm 6.5cm 22cm},clip]{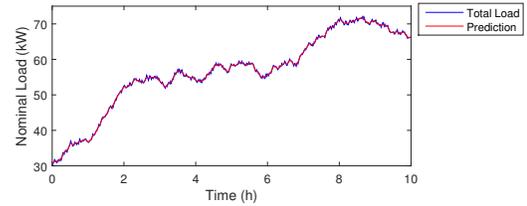}
\caption{\textcolor{Black}{Nominal total microgrid load profile (for bus voltages of 415V) and predictions.}}
\label{fig:Load}
\end{figure}

\begin{figure}[!t]
\centering
\includegraphics[width=0.8\columnwidth,trim={0.0cm 0cm 6.5cm 22cm}]{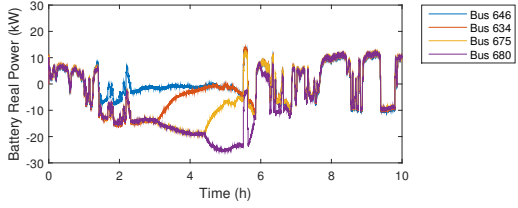}
\caption{Battery ES system real output powers.}
\label{fig:Pbatt}
\end{figure}

\begin{figure}[!t]
\centering
\includegraphics[width=0.8\columnwidth,trim={0.0cm 0cm 6.5cm 22cm}]{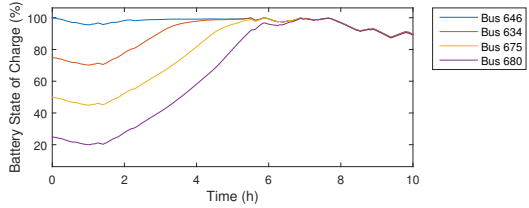}
\caption{Battery ES system SoC levels.}
\label{fig:SoC}
\end{figure}

\begin{figure}[!t]
\centering
\includegraphics[width=0.8\columnwidth,trim={0.0cm 0cm 6.5cm 22cm}]{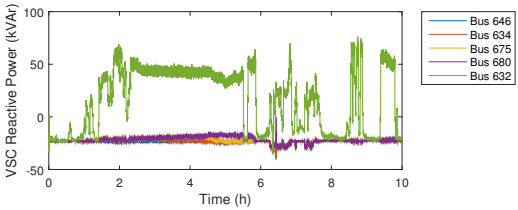}
\caption{VSC reactive output powers.}
\label{fig:Q}
\end{figure}

\begin{figure}[!t]
\centering
\includegraphics[width=0.8\columnwidth,trim={0.0cm 0cm 6.5cm 22cm}]{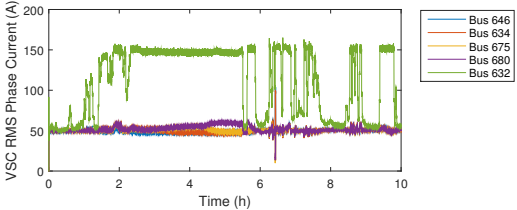}
\caption{VSC RMS inductor phase currents.}
\label{fig:Irms}
\end{figure}

\begin{figure}[!t]
\centering
\includegraphics[width=0.8\columnwidth,trim={0.0cm 0cm 6.5cm 22cm}]{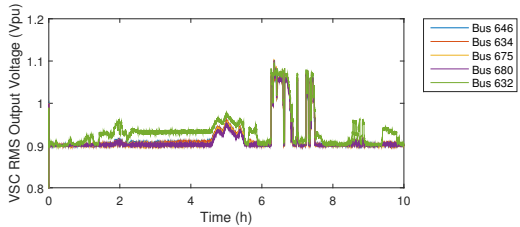}
\caption{VSC per unit RMS output voltages.}
\label{fig:Vrms}
\end{figure}

The MPC is formulated with a 1 minute sampling period based on the PV generation variability. A 30 minute time horizon is used based on the expected availability of PV generation predictions from ground-based sky imaging sensors \cite{Diagne2013}. \textcolor{Black}{At each sampling interval, the MPC strategy has access to the current SoC of the batteries, PV generation and microgrid load, as well as inaccurate PV generation and load predictions over the time horizon.}

\textcolor{Black}{Fig. \ref{fig:Ppv} shows the varying PV generation output power and the predictions used by the MPC strategy. The PV generation predictions are modelled by a moving average with a 5 minute window. This gives a root mean squared error of 12.7\%, which is in line with the error expected from intra-hour PV generation predictions \cite{doi:10.1175/JTECH-D-13-00209.1}}. 

\textcolor{Black}{The microgrid has a variable resistive load at each bus. Fig. \ref{fig:Load} shows the total nominal load profile (for bus voltages of 415V) and the load predictions, which are modelled by a moving average with a 5 minute window. The total load is evenly divided between the 12 microgrid buses. The load profile was generated based on the residential microgrid load profile from \cite{Papathanassiou2005}. A random walk (with uniformly distributed steps between $\pm0.75\%$ each minute) was added to model minute to minute variations.} 

The MPC QCQP was solved in MATLAB with the IBM CPLEX solver, implemented using the YALMIP toolbox \cite{YALMIP}. The average solution time was 0.71 seconds on an Intel Core i7-4770 CPU. TCP/IP communication over an Ethernet network was used between the computer running the MPC and the real-time digital simulator simulating the network and local VSC controllers. 

Fig. \ref{fig:SoC} shows that the battery ES systems begin with SoC between 25\% and 100\%. As shown in Fig. \ref{fig:Pbatt}, the battery ES systems share the difference between the intermittent PV generation and the load. \textcolor{Black}{The minimum SoC reached by any of the batteries is 19.98\% and the maximum SoC reached is 100.05\%, slightly violating the 20\% to 100\% limits. These small violations are caused by the approximate nature of the battery SoC model.}

The VSCs inject/absorb reactive power to control the microgrid bus voltages, as shown in Fig. \ref{fig:Q}. Fig. \ref{fig:Irms} shows the VSC phase currents. The maximum phase current is 165A. \textcolor{Black}{Averaging over each 1 minute sampling period to remove the ripple current and transients (which are not considered by the MPC strategy) the maximum average phase current for any interval is 154.8A, slightly violating the 150A limit. The constraint violation is caused by the PV generation prediction errors.} Fig. \ref{fig:Vrms} shows that the microgrid bus voltages are kept within the required limits of $V_{LL}\pm$10\%. \textcolor{Black}{The proposed MPC strategy keeps the VSC output voltages near the lower voltage limit. Since the microgrid load is mainly resistive, this reduces the load currents and losses. This demonstrates the importance of an MPC strategy that includes the effect of line impedances and source output voltages on losses.}

\textcolor{Black}{The average power loss is 12.638kW over the case study. To ensure the proposed MPC strategy is not overly sensitive to the placement of the VSCs, the case study was repeated with the PV generation source at bus 611, bus 652 and bus 645 (instead of bus 632), giving average losses of 12.634kW, 12.642kW and 12.654kW respectively.}

\subsection{Comparison with the Non-Convex QCQP}

The accuracy of the sub-optimal solution found by the proposed MPC strategy based on the convex QCQP formulation  (\ref{equ:convexQP}) with inaccurate PV generation and load predictions, was compared to the non-convex QCQP formulation (\ref{equ:nonconvexQP}) without PV generation and load prediction errors. \textcolor{Black}{Solving the non-convex QCQP takes too long for real-time implementation. Instead it was solved offline for each thirty minute interval of the case study, based on accurate PV generation and load predictions. MATLAB's FMINCON solver was used to find a local optimum, taking the solution of the convex QCQP as a starting point.}

\textcolor{Black}{The convex QCQP gave an average power loss of 12.638kW over the case study, while the non-convex QCQP gave an average power loss of 11.596kW (8.25\% lower). However, the average solution time for the non-convex QCQP was 23 minutes, compared to 0.71 seconds for the convex QCQP (a reduction by a factor of 1000), demonstrating its unsuitability for a real-time implementation.}

\subsection{\textcolor{Black}{Comparison with a Constant Efficiency SoC Model}}

\textcolor{Black}{To quantify the benefits of the proposed variable efficiency battery SoC model, the proposed MPC strategy formulation was modified so that the battery charging and discharging efficiencies were kept constant. A charging efficiency of 99.81\% and discharging efficiency of 99.80\% were used based on a nominal charging and discharging power of $\pm$5kW and a SoC of 70\%. The case study was repeated using the constant efficiency MPC strategy, giving average losses of 13.194kW, a 4.40\% higher cost than the variable efficiency formulation (12.638kW).}

\textcolor{Black}{The proposed method for incorporating the variable battery efficiency is approximate and, in certain cases, it may be possible to find fixed efficiency values which give superior performance. The main advantage of the proposed method is that it provides a means of generating reasonable approximate values based on the microgrid operating state.}

\section{Conclusion}
\label{sec:Conclusion}
\textcolor{Black}{A new convex MPC strategy for dynamic optimal power flow between distributed AC microgrid battery ES systems has been presented. The convex optimisation formulation can be solved quickly using robust solvers, and accounts for line losses and voltage drops in the microgrid. The proposed control strategy approaches the performance of a strategy based on non-convex optimisation, while reducing the required computation time by a factor of 1000, making it suitable for a real-time MPC implementation.}




\ifCLASSOPTIONcaptionsoff
  \newpage
\fi

\bibliographystyle{IEEEtran}  
\bibliography{IEEEabrv,ES_MPC_Bibtex}

\begin{thebibliography}{10}
\providecommand{\url}[1]{#1}
\csname url@samestyle\endcsname
\providecommand{\newblock}{\relax}
\providecommand{\bibinfo}[2]{#2}
\providecommand{\BIBentrySTDinterwordspacing}{\spaceskip=0pt\relax}
\providecommand{\BIBentryALTinterwordstretchfactor}{4}
\providecommand{\BIBentryALTinterwordspacing}{\spaceskip=\fontdimen2\font plus
\BIBentryALTinterwordstretchfactor\fontdimen3\font minus
  \fontdimen4\font\relax}
\providecommand{\BIBforeignlanguage}[2]{{%
\expandafter\ifx\csname l@#1\endcsname\relax
\typeout{** WARNING: IEEEtran.bst: No hyphenation pattern has been}%
\typeout{** loaded for the language `#1'. Using the pattern for}%
\typeout{** the default language instead.}%
\else
\language=\csname l@#1\endcsname
\fi
#2}}
\providecommand{\BIBdecl}{\relax}
\BIBdecl

\bibitem{Nehrir2011}
M.~H. Nehrir, C.~Wang, K.~Strunz, H.~Aki, R.~Ramakumar, J.~Bing, Z.~Miao, and
  Z.~Salameh, ``{A Review of Hybrid Renewable/Alternative Energy Systems for
  Electric Power Generation: Configurations, Control, and Applications},''
  \emph{{IEEE} Trans. Sustain. Energy}, vol.~2, no.~4, pp. 392--403, Oct. 2011.

\bibitem{Lasseter2002}
R.~Lasseter, ``{MicroGrids},'' in \emph{2002 IEEE Power Engineering Society
  Winter Meeting.}, vol.~1, 2002, pp. 305--308.

\bibitem{Momoh1999}
J.~Momoh, R.~Adapa, and M.~El-Hawary, ``{A review of selected optimal power
  flow literature to 1993. I. Nonlinear and quadratic programming
  approaches},'' \emph{{IEEE} Trans. Power Syst.}, vol.~14, no.~1, pp. 96--104,
  1999.

\bibitem{Levron2013}
Y.~Levron, J.~M. Guerrero, and Y.~Beck, ``{Optimal Power Flow in Microgrids
  With Energy Storage},'' \emph{{IEEE} Trans. Power Syst.}, vol.~28, no.~3, pp.
  3226--3234, Aug. 2013.

\bibitem{Agarwal2010}
V.~Agarwal, K.~Uthaichana, R.~A. DeCarlo, and L.~H. Tsoukalas, ``{Development
  and Validation of a Battery Model Useful for Discharging and Charging Power
  Control and Lifetime Estimation},'' \emph{{IEEE} Trans. Energy Convers.},
  vol.~25, no.~3, pp. 821--835, Sep. 2010.

\bibitem{Chen2006}
M.~Chen and G.~Rincon-Mora, ``{Accurate Electrical Battery Model Capable of
  Predicting Runtime and I–V Performance},'' \emph{{IEEE} Trans. Energy
  Convers.}, vol.~21, no.~2, pp. 504--511, Jun. 2006.

\bibitem{Parvini2015}
Y.~Parvini and A.~Vahidi, ``{Maximizing charging efficiency of lithium-ion and
  lead-acid batteries using optimal control theory},'' \emph{2015 American
  Control Conference (ACC)}, pp. 317--322, Jul. 2015.

\bibitem{Olivares2014}
D.~E. Olivares, A.~Mehrizi-Sani, A.~H. Etemadi, C.~A. Canizares, R.~Iravani,
  M.~Kazerani, A.~H. Hajimiragha, O.~Gomis-Bellmunt, M.~Saeedifard,
  R.~Palma-Behnke, G.~A. Jimenez-Estevez, and N.~D. Hatziargyriou, ``{Trends in
  Microgrid Control},'' \emph{{IEEE} Trans. Smart Grid}, vol.~5, no.~4, pp.
  1905--1919, Jul. 2014.

\bibitem{Christofides2013}
P.~D. Christofides, R.~Scattolini, D.~{Mu\~{n}oz de la Pe\~{n}a}, and J.~Liu,
  ``{Distributed model predictive control: A tutorial review and future
  research directions},'' \emph{Computers \& Chemical Engineering}, vol.~51,
  pp. 21--41, Apr. 2013.

\bibitem{Lee2011}
J.~H. Lee, ``{Model predictive control: Review of the three decades of
  development},'' \emph{International Journal of Control, Automation and
  Systems}, vol.~9, no.~3, pp. 415--424, Jun. 2011.

\bibitem{Parisio2014}
A.~Parisio, E.~Rikos, and L.~Glielmo, ``{A Model Predictive Control Approach to
  Microgrid Operation Optimization},'' \emph{{IEEE} Trans. Control Syst.
  Technol.}, vol.~22, no.~5, pp. 1813--1827, Sep. 2014.

\bibitem{Perez2013}
E.~Perez, H.~Beltran, N.~Aparicio, and P.~Rodriguez, ``{Predictive Power
  Control for PV Plants With Energy Storage},'' \emph{{IEEE} Trans. Sustain.
  Energy}, vol.~4, no.~2, pp. 482--490, Apr. 2013.

\bibitem{Mayhorn2012}
E.~Mayhorn, K.~Kalsi, M.~Elizondo, N.~Samaan, and K.~Butler-Purry, ``{Optimal
  control of distributed energy resources using model predictive control},'' in
  \emph{2012 IEEE Power and Energy Society General Meeting}, Jul. 2012, pp.
  1--8.

\bibitem{Dagdougui2014}
H.~Dagdougui, L.~Dessaint, and A.~Ouammi, ``{Optimal power exchanges in an
  interconnected power microgrids based on model predictive control},''
  \emph{2014 IEEE PES General Meeting | Conference \& Exposition}, pp. 1--5,
  Jul. 2014.

\bibitem{Garcia-Torres2015}
F.~Garcia-Torres and C.~Bordons, ``{Optimal Economical Schedule of
  Hydrogen-Based Microgrids With Hybrid Storage Using Model Predictive
  Control},'' \emph{{IEEE} Trans. Ind. Electron.}, vol.~62, no.~8, pp.
  5195--5207, Aug. 2015.

\bibitem{Zeng2014}
P.~P. Zeng, Z.~Wu, X.-P. Zhang, C.~Liang, and Y.~Zhang, ``{Model predictive
  control for energy storage systems in a network with high penetration of
  renewable energy and limited export capacity},'' \emph{2014 Power Systems
  Computation Conference}, pp. 1--7, Aug. 2014.

\bibitem{Tabatabaee2011}
S.~Tabatabaee, H.~R. Karshenas, A.~Bakhshai, and P.~Jain, ``{Investigation of
  droop characteristics and X/R ratio on small-signal stability of autonomous
  Microgrid},'' \emph{2011 2nd Power Electronics, Drive Systems and
  Technologies Conference}, pp. 223--228, Feb. 2011.

\bibitem{Olivares2014b}
D.~E. Olivares, C.~A. Canizares, and M.~Kazerani, ``{A Centralized Energy
  Management System for Isolated Microgrids},'' \emph{{IEEE} Trans. Smart
  Grid}, vol.~5, no.~4, pp. 1864--1875, Jul. 2014.

\bibitem{Pogaku2007}
N.~Pogaku, M.~Prodanovic, and T.~C. Green, ``{Modeling, Analysis and Testing of
  Autonomous Operation of an Inverter-Based Microgrid},'' \emph{{IEEE} Trans.
  Power Electron.}, vol.~22, no.~2, pp. 613--625, Mar. 2007.

\bibitem{Savaghebi2012}
M.~Savaghebi, A.~Jalilian, J.~C. Vasquez, and J.~M. Guerrero, ``{Secondary
  Control Scheme for Voltage Unbalance Compensation in an Islanded
  Droop-Controlled Microgrid},'' \emph{{IEEE} Trans. Smart Grid}, vol.~3,
  no.~2, pp. 797--807, jun 2012.

\bibitem{Ciobotaru2007}
M.~Ciobotaru, R.~Teodorescu, P.~Rodriguez, A.~Timbus, and F.~Blaabjerg,
  ``{Online grid impedance estimation for single-phase grid-connected systems
  using PQ variations},'' \emph{PESC Record - IEEE Annual Power Electronics
  Specialists Conference}, pp. 2306--2312, Jun. 2007.

\bibitem{Wu2007}
X.~Wu, S.~K. Panda, and J.~Xu, ``{Effect of Pulse-Width Modulation Schemes on
  the Performance of Three-Phase Voltage Source Converter},'' \emph{IECON 2007
  - 33rd Annual Conference of the IEEE Industrial Electronics Society}, pp.
  2026--2031, 2007.

\bibitem{Kim2011}
T.~Kim and W.~Qiao, ``{A Hybrid Battery Model Capable of Capturing Dynamic
  Circuit Characteristics and Nonlinear Capacity Effects},'' \emph{{IEEE}
  Trans. Energy Convers.}, vol.~26, no.~4, pp. 1172--1180, Dec. 2011.

\bibitem{Johansson2004}
S.~G. Johansson, G.~Asplund, E.~Jansson, and R.~Rudervall, ``{Power system
  stability benefits with VSC DC-transmission systems.}'' in \emph{CIGRE
  Conference}, Paris, France, 2004, pp. 1--8.

\bibitem{Boyd2004}
S.~Boyd and L.~Vandenberghe, \emph{{Convex Optimization}}.\hskip 1em plus 0.5em
  minus 0.4em\relax Cambridge University Press, 2004.

\bibitem{Villalva2009}
M.~Villalva, J.~Gazoli, and E.~Filho, ``{Comprehensive Approach to Modeling and
  Simulation of Photovoltaic Arrays},'' \emph{{IEEE} Trans. Power Electron.},
  vol.~24, no.~5, pp. 1198--1208, May 2009.

\bibitem{Diagne2013}
M.~Diagne, M.~David, P.~Lauret, J.~Boland, and N.~Schmutz, ``{Review of solar
  irradiance forecasting methods and a proposition for small-scale insular
  grids},'' \emph{Renewable and Sustainable Energy Reviews}, vol.~27, pp.
  65--76, Nov. 2013.

\bibitem{doi:10.1175/JTECH-D-13-00209.1}
Y.~Chu, H.~T.~C. Pedro, L.~Nonnenmacher, R.~H. Inman, Z.~Liao, and C.~F.~M.
  Coimbra, ``A smart image-based cloud detection system for intrahour solar
  irradiance forecasts,'' \emph{Journal of Atmospheric and Oceanic Technology},
  vol.~31, no.~9, pp. 1995--2007, 2014.

\bibitem{Papathanassiou2005}
S.~Papathanassiou, N.~Hatziargyriou, and K.~Strunz, ``{A Benchmark Low Voltage
  Microgrid Network},'' in \emph{Proceedings of the CIGRE Symposium: Power
  Systems with Dispersed Generation}, Jan. 2005, pp. 1--8.

\bibitem{YALMIP}
J.~Löfberg, ``Yalmip : A toolbox for modeling and optimization in {MATLAB},''
  in \emph{Proceedings of the CACSD Conference}, Taipei, Taiwan, 2004.

\end{thebibliography}

\end{document}